\documentclass{article}
\usepackage{tabularx}
\usepackage{caption}
\usepackage{booktabs}
\usepackage{spconf,amsmath,graphicx}
\usepackage{float}

\title{EmoHRNet: HIGH-RESOLUTION NEURAL
NETWORK Based SPEECH EMOTION RECOGNITION}
\name{Akshay Muppidi, Martin Radfar}
\address{Stony Brook University \\ Department of Computer Science\\ Stony Brook, New York, USA}
\begin{document}
%
\maketitle
\begin{abstract}
Speech emotion recognition (SER) is pivotal for enhancing human-machine interactions. This paper introduces "EmoHRNet", a novel adaptation of High-Resolution Networks (HRNet) tailored for SER. The HRNet structure is designed to maintain high-resolution representations from the initial to the final layers. By transforming audio samples into spectrograms, EmoHRNet leverages the HRNet architecture to extract high-level features. EmoHRNet's unique architecture maintains high-resolution representations throughout, capturing both granular and overarching emotional cues from speech signals. The model outperforms leading models, achieving accuracies of 92.45\% on RAVDESS, 80.06\% on IEMOCAP, and 92.77\% on EMOVO. Thus, we show that EmoHRNet sets a new benchmark in the SER domain.
\end{abstract}

\section{Introduction}
\label{sec:intro}
Speech emotion recognition (SER) has emerged as a pivotal domain, instrumental in advancing robot intelligence and human-machine interactions \cite{article}. Recognizing emotions from speech signals can substantially enhance the communication quality between humans and machines. However, discerning emotions from speech signals remains intricate due to factors like background noise, individual-specific accentuation, weak representation of grammatical and semantic knowledge, and the unique temporal and spectral attributes of speech signals \cite{7975169}.

Recent literature has spotlighted the potential of High-Resolution Networks (HRNet) for tasks demanding high-resolution inputs, especially in image analysis \cite{9052469}. HRNet's design, with its multi-resolution strategy that simultaneously extracts features from varying scales, allows it to assimilate both granular and overarching information, offering an edge in accuracy and speed over other models \cite{sun2019deep}. In this context, we introduce "EmoHRNet", a novel adaptation of HRNet tailored for SER. We transform audio samples into spectrograms and employ the HRNet architecture to glean high-level features from these visual representations. Moreover, we use data augmentation techniques to capitalize on the intrinsic link between emotions in speech and variations in pitch, tone, and temporal patterns. Our experimental findings underscore that the HRNet-based SER model  surpasses other leading models in unweighted accuracy. Specifically, our model achieves unweighted accuracies of 92.45\% on RAVDESS, 80.06\% on IEMOCAP, and 92.77\% on EMOVO. 

\section{Relation to Prior Work}
\label{sec:format}

A myriad of techniques have been proposed to tackle the challenges of SER. With the advent of deep learning, newer models like deep neural networks combined with extreme learning machines \cite{AKYOL2020112875}, bi-directional Long Short-Term Memory (LSTM) \cite{SENTHILKUMAR20222180}, Recurrent Neural Networks (RNN) \cite{inproceedings},  Capsule Neural Networks \cite{9412360} \cite{8683163}, and Quaternion based CNNs \cite{9414248} have shown promise in capturing high-level representations from pitch-based features and other speech attributes.

Attention-based SER models, such as those employing multi-head attention \cite{9054073} and attention pooling \cite{Li2018AnAP}, have been increasingly studied for their potential in extracting high-level emotional information. However, many of these models, despite their advanced capabilities, are often laden with a large number of parameters, making them less suitable for real-time applications and environments constrained by computational resources.

Furthermore, while models like the dual-level LSTM \cite{Wang_2020}, which harnesses temporal information from different time-frequency resolutions, and the integrated spatiotemporal feature learners \cite{LI2021238}, have shown potential, they often face challenges. One of the primary limitations is their inability to consistently capture long-range dependencies essential for context modeling in SER. Emotions in speech are intrinsically context-dependent, and a model's failure to grasp these dependencies can lead to inaccuracies. Additionally, many of these models do not dynamically adjust their receptive fields, which can limit their adaptability and generalization to unfamiliar data or diverse corpora. While recent advancements like the Capsule neural network-based CNN \cite{wen2022ctlmtnet}, the Gated multi-scale temporal convolutional network \cite{Ye_2022}, temporal modeling \cite{10096370}, and multi-resolution feature extraction methods \cite{10095260} have shown potential, there remains a gap in consistently achieving high accuracies across diverse datasets and real-world scenarios.

In light of these challenges and limitations, High-Resolution Networks (HRNet) emerges as a promising solution. HRNet's unique architecture, which maintains high-resolution representations through parallel multi-resolution convolutions, allows it to capture both fine-grained and coarse contextual information simultaneously. This multi-resolution strategy is particularly advantageous for SER, where capturing nuances at different scales is crucial. Unlike many models that downsample and then upsample, HRNet's consistent high-resolution processing ensures that no critical emotional cues are lost. Moreover, its design inherently addresses the limitation of models that struggle with long-range dependencies, as HRNet can assimilate both granular and overarching information seamlessly. Our adaptation of HRNet, "EmoHRNet", further tailors this architecture for SER, achieving superior performance metrics across benchmark datasets.  To the best of our knowledge, this is the first time that HRNet is being applied to the domain of SER. Notably, EmoHRNet outperforms the aforementioned state-of-the-art methods in accuracy, including attention-based models, making it an optimal choice for real-world SER applications.

\section{Model}
\subsection{Preprocessing and Data Augmentation}

\begin{figure}[H]
  \centering
  \includegraphics[width=8.9cm, height=3.243cm]{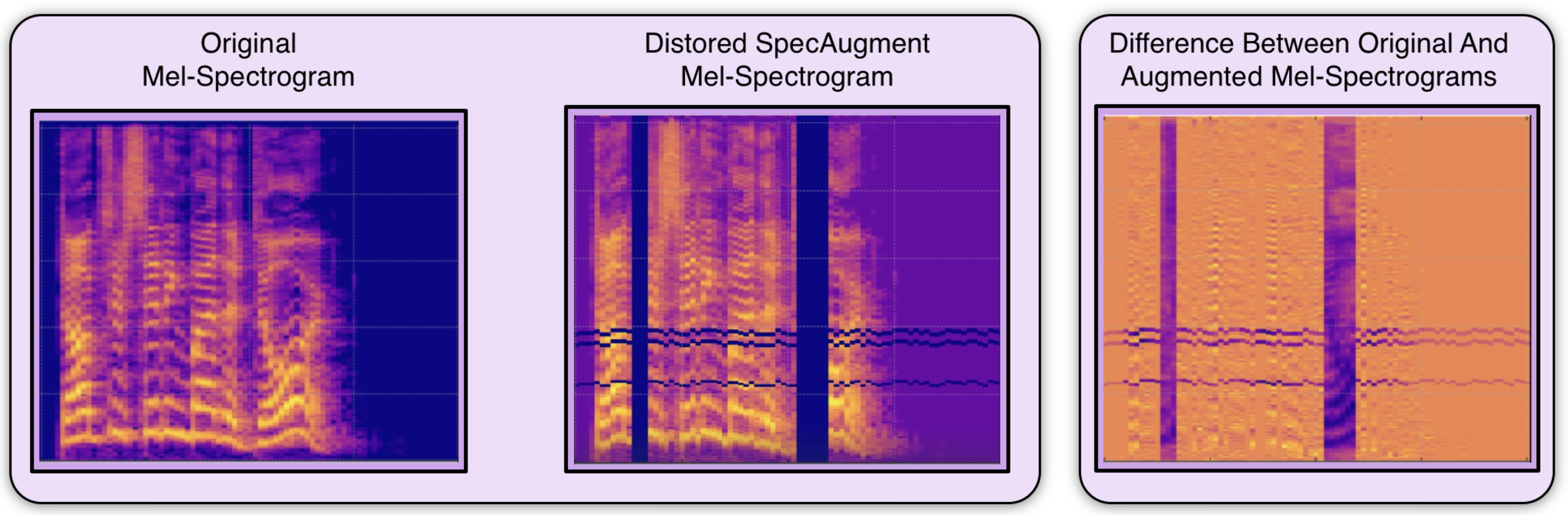}
  \caption{{The Original Mel-Spectrogram, The Distorted SpecAugment Mel-Spectrogram, and The Difference Between Orginal and Augmented Mel-Spectrograms. }}
  \label{fig1}
\end{figure}

Audio signals are transformed into Mel-spectrograms using STFT[15] and are normalized. For augmentation, Mel-spectrograms are randomly shifted along the time axis. Moreover, we use a commonly used augmentation technique: SpecAugment \cite{Park_2019}, specifically frequency masking and time masking. Frequency masking obscures frequency bands, chosen with \(f \sim U(0, F)\), based on the distribution of pitch variations in the training set. Time masking masks consecutive time steps, defined by \(t \sim U(0, T)\), set according to typical emotional utterance durations. Refer to \textbf{Fig \ref{fig1}} for a visual representation of augmented data.

\subsection{HRNet Structure}

The HRNet architecture is meticulously designed to maintain high-resolution representations from the initial to the final layers. This consistent high-resolution processing is crucial for tasks like SER, where the detailed nuances in Mel spectrogram inputs are essential for accurate emotion recognition.\\

 \textbf{High-Resolution Input Module (HRIM):} At the outset, the HRIM processes the Mel spectrogram. It employs a 3x3 convolution to extract preliminary features, setting the stage for the deeper layers of the network. This initial processing ensures that the network starts with a rich set of features derived from the input.
    
 \textbf{High-Resolution Stages (HRS):} As the architecture deepens, it doesn't compromise on resolution. Instead, it introduces parallel branches that operate at varying resolutions. These branches are not isolated; they exchange information through a mechanism that allows multi-resolution fusions. This design ensures that the network captures and integrates features across multiple scales, preserving both granular details and broader patterns.
    
\textbf{Fuse Layer (FL):} Serving as a unifying layer, the FL takes the multi-resolution feature maps from the various stages and fuses them. It employs 1x1 convolutions to consolidate these maps into a singular high-resolution feature map. This fusion process ensures that the final output is a comprehensive representation that has benefited from multi-scale processing. To counteract potential challenges like the vanishing gradient problem inherent in deep networks, residual connections are strategically placed throughout the network.

\begin{figure*}
  \includegraphics[width=\textwidth,height=10cm]{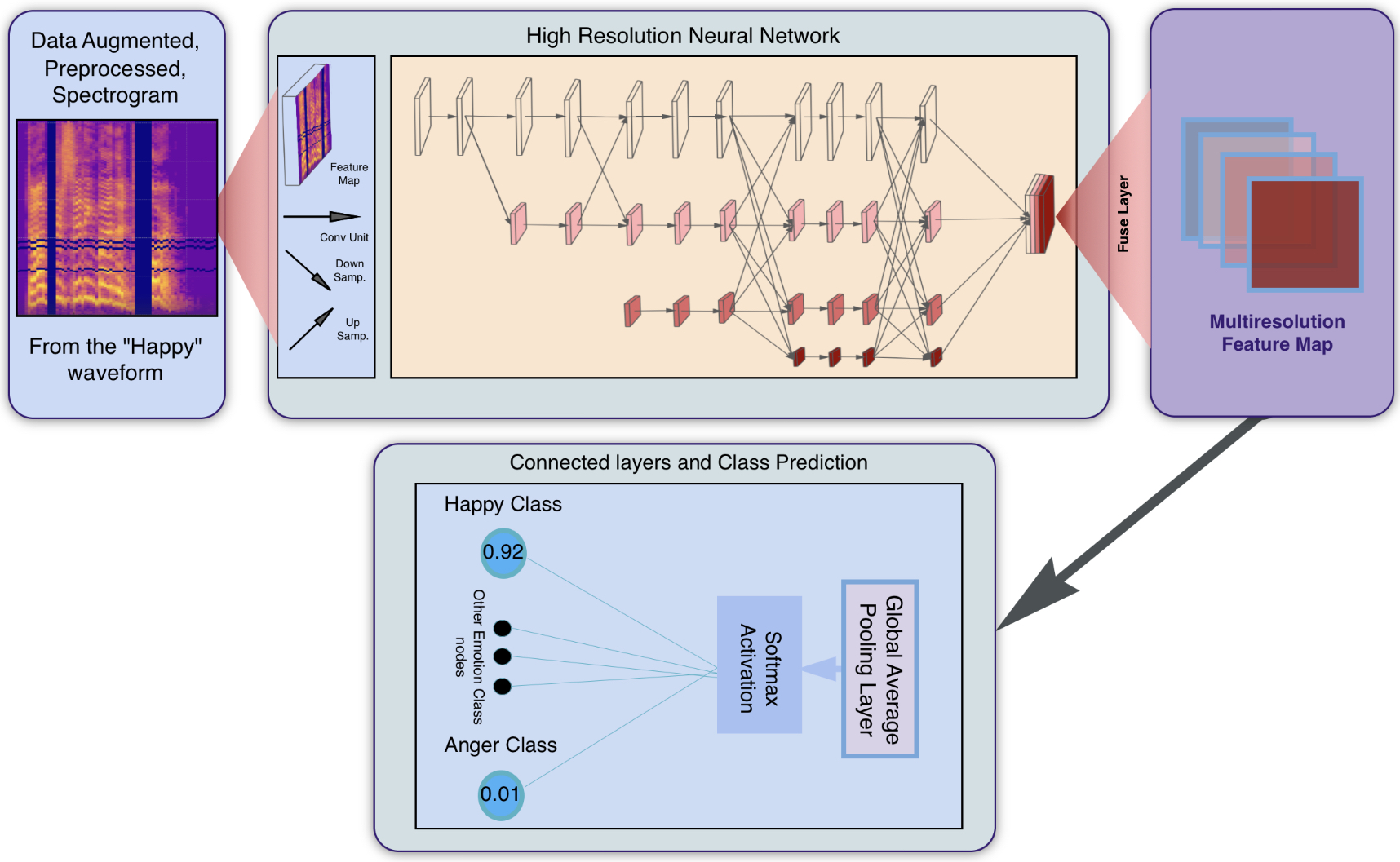}
  \caption{{EmoHRNet Model Architecture: Input, High Resolution Stages, Fuse Layer, and Fully Connected Layers.}}
  \label{fig2}
\end{figure*}
\begin{table*}[h]
\centering

\begin{tabular}{l|c|c|l|c|c|l|c|c}
\toprule
\multicolumn{3}{c|}{IEMOCAP} & \multicolumn{3}{c|}{RAVDESS} & \multicolumn{3}{c}{EMOVO} \\
\cmidrule(lr){1-3} \cmidrule(lr){4-6} \cmidrule(lr){7-9}
Model & Year & Accuracy & Model & Year & Accuracy & Model & Year & Accuracy \\
\midrule
Zhong et al & 2020 & 71.72\% & QCNN & 2021 & 77.87\% & Tuncer et al & 2021 & 79.08\% \\
QCNN & 2021 & 70.46\% & CTL-MTNet & 2022 & 90.83\% & CTL-MTNet & 2022 & 85.40\% \\

Light-SERNet & 2021 & 70.78\% & Hybrid MFCCT + CNN & 2023 & 92.00\% & Al-onazi et al & 2022 & 91.70\% \\
ACNN+SE & 2022 & 75.00\% & ACNN+SE & 2022 & 78.77\% & Xie et al & 2023 & 89.24\% \\
TIM-Net & 2023 & 71.65\% & TIM-Net & 2023 & 92.08\% & TIM-Net & 2023 & 92.00\% \\
TWATWF + BCNN & 2023 & 79.07\% & TWATWF + BCNN & 2023 & 80.37\% & Sekkate et al & 2023 & 83.90\% \\
\textbf{EmoHRNet} & \textbf{2024} & \textbf{80.06\%} & \textbf{EmoHRNet} & \textbf{2024} & \textbf{92.45\%} & \textbf{EmoHRNet} & \textbf{2024} & \textbf{92.77\%} \\
\bottomrule
\end{tabular}
\caption{Results on EmoHRNet and state-of-the-art models for IEMOCAP, RAVDESS, and EMOVO}
\label{table:results_data}

\end{table*}

\subsection{Connecting Layers}
The output multiresolution feature map $F_{FL}$ from the Fuse Layer (FL) is directed to the connecting layers for the classification task. These layers comprise a global average pooling layer[19], which averages each feature map across its spatial dimensions, producing a fixed-size feature vector. This vector is then passed to a fully connected layer, which employs a softmax activation function[8] to generate a probability distribution over the emotion classes. The output from this layer is represented as $y$, given by:

\begin{equation}
z_i = \frac{1}{HW}\sum_{h=1}^{H}\sum_{w=1}^{W}F_{FL,i,h,w}
\end{equation}
\begin{equation}
y_i = \frac{e^{z_i}}{\sum_{j=1}^{C}e^{z_j}}
\end{equation}
here, $C$ denotes the number of emotion classes, $H$ and $W$ represent the spatial dimensions of the feature map, and $F_{FL,i,h,w}$ is the activation of the $i$th channel at spatial location $(h, w)$ in the feature map $F_{FL}$. The full architecture of the proposed EmoHRNet model is visualized in \textbf{Fig \ref{fig2}}.

\subsection{Training}
The proposed HRNet-based SER model is trained using the cross-entropy loss function, which measures the difference between the predicted probabilities and the ground-truth labels for each sample:
\begin{equation}
L = -\frac{1}{N}\sum_{i=1}^N \sum_{c=1}^C y_{i,c} \log(p_{i,c})
\end{equation}

In this equation, $N$ stands for the number of training samples, $C$ is the number of emotion classes, $y_{i,c}$ is the true label of the $i$th sample for the $c$th emotion class, and $p_{i,c}$ is the model's predicted probability for the same.

For optimization, we employ the Adam optimizer with parameters: learning rate set to 0.001, beta1 at 0.9, and beta2 at 0.999. To mitigate overfitting, weight decay regularization is applied with a coefficient of 0.0001. The model is trained over 100 epochs with batches of 64 samples each. Model performance is periodically assessed on a validation set, and the iteration with the highest validation accuracy is selected as the final model.

\section{Experiments}

\subsection{Materials}

This study employs three benchmarked datasets for speech emotion recognition: RAVDESS\cite{AKYOL2020112875}, IEMOCAP\cite{IEMOCAP}, and EMOVO\cite{costantini-etal-2014-emovo}.

\subsubsection{RAVDESS}
The Ryerson Audio-Visual Database of Emotional Speech and Song (RAVDESS) comprises 7356 audio files from 24 professional actors, covering eight emotions in both speech and song formats. Each emotion is represented in two intensities: normal and strong.

\subsubsection{IEMOCAP}
The Interactive Emotional Dyadic Motion Capture Database (IEMOCAP) offers 12 hours of audiovisual interactions between actors, capturing emotions like happiness, anger, sadness, frustration, and neutral. 

\subsubsection{EMOVO}
EMOVO is a pioneering emotional corpus tailored for the Italian language. It comprises recordings from six actors who articulated 14 sentences, capturing disgust, fear, anger, joy, surprise, and sadness, in addition to a neutral state. The corpus underwent a validation process with two distinct groups of 24 listeners, achieving an 80\% recognition accuracy.

\subsection{Results}

We assessed the performance of our proposed EmoHRNet model on three renowned speech emotion recognition datasets: IEMOCAP, RAVDESS, and EMOVO. The results were juxtaposed with those of previously published state-of-the-art models, as shown in Table 1. Notably, we compared the following state-of-the-art models: Separable Convolution\cite{Zhong2020}, QCNN\cite{9414248} , Light-SERNet \cite{9746679}, ACNN+SE \cite{inbook}, Tuncer et al\cite{tuncer2021automated}, TIM-Net \cite{10096370}, TWATWF + BCNN \cite{10095260}, CTL-MTNet \cite{wen2022ctlmtnet}, Hybrid MFCCT + CNN \cite{Alluhaidan}, Transformer with Feature Fusion \cite{Al-onazi}, Two-Stage feature selection \cite{XIE2023102955}, and statistical feature extraction \cite{Sekkate2023}.

From Table \ref{table:results_data}, it is evident that EmoHRNet consistently outperforms other leading models across all datasets. Specifically, on the RAVDESS dataset, EmoHRNet achieved an accuracy of 92.45\%, for the IEMOCAP dataset, EmoHRNet's accuracy of 80.06\% stands out, and for the EMOVO dataset, it achieves an accuracy of 92.77\%.

The superior performance of EmoHRNet can be attributed to several factors. Primarily, the HRNet architecture's ability to maintain high-resolution representations throughout its depth allows for the extraction and preservation of intricate emotional features from the speech spectrograms. An interesting discussion point is that while the TWATWF + BCNN model employs a multi-branch network structure to capture features across different time and frequency dimensions, EmoHRNet offers a more holistic structured method. By seamlessly integrating multi-resolution features in a hierarchical manner, EmoHRNet ensures robust and adaptive feature extraction, guaranteeing resilience and high performance across diverse scenarios. This may be why it performed similarly, but still better than, TWATWF + BCNN. 

The results underscore the efficacy of EmoHRNet in speech emotion recognition tasks, setting a new benchmark for future research in this domain. 

\section{Conclusion}

In this paper, we introduced EmoHRNet, a novel model for speech emotion recognition (SER) that leverages the strengths of the HRNet architecture. Our approach emphasizes the importance of maintaining high-resolution representations throughout the network's depth, ensuring the extraction and preservation of intricate emotional features from speech spectrograms. The results, as demonstrated on three renowned SER datasets— IEMOCAP, RAVDESS, and EMOVO—highlight the model's superior performance, setting a new benchmark in the domain.

Future research could delve into the selection of different features, particularly focusing on the extraction of prosodic, phonetic, and articulatory features, which have been shown to carry significant emotional information. Combining EmoHRNet with other models and methods discussed in this paper could potentially lead to even more robust and accurate SER systems. Moreover, experimenting with other data augmentation techniques, beyond the ones employed in this study, might further improve the model's generalization.

\bibliographystyle{IEEEbib}
\small{
\bibliography{draft1Ref}
}
\footnotesize
\end{document}